\documentclass[aps,twocolumn,showpacs]{revtex4}
\usepackage{graphics}
\usepackage{graphicx}

\def\ss{\scriptscriptstyle }
\begin{document}
\title{Transient response of photoexcited electrons: \\
negative and oscillating current}
\author{O.E. Raichev}
\author{F.T. Vasko}
\email{ftvasko@yahoo.com}
\affiliation{Institute of Semiconductor Physics, National Academy of Sciences of Ukraine,
03028, Kiev, Ukraine}
\date{\today}

\begin{abstract}
Time-dependent current of the electrons excited in the conduction band after ultrafast
interband photogeneration is studied theoretically. The transient photocurrent is
calculated for the nonlinear regime of response to a stationary electric field.
The response demonstrates transient absolute negative conductivity when the electrons
are excited slightly below the optical phonon energy, while the periodic oscillations
of the electric current appear after formation of the streaming distribution. The
quenching of these peculiarities by the elastic scattering of electrons is also
considered.
\end{abstract}


\maketitle

The steady-state nonlinear response of electrons with a strongly anisotropic
(streaming) distribution appears due to the cyclic motion of electrons
accelerating in the passive region ($\varepsilon < \hbar \omega_o$) and rapidly
emitting optical phonons when penetrating into the active region ($\varepsilon >
\hbar \omega_o$). This transport regime takes place under the conditions
$\nu \ll t_{\ss E}^{-1} \ll \nu_{o}$, see Sec. 14.2 in Ref. 1 and Sec. 35 in Ref. 2.
Here and below, $t_{\ss E}=p_o/|e|E$ is the time of ballistic flight of an electron across the
passive region, $E$ the strength of the applied electric field, $e$ the electron charge,
$\nu$ the elastic scattering rate, $\nu_o$ the rate of spontaneous emission of optical
phonons, $p_o \equiv \sqrt{2m \hbar \omega_o}$ the electron momentum corresponding
to the optical phonon energy $\hbar \omega_o$, and $m$ the electron effective mass.
Stimulated THz emission of bulk holes in the streaming regime remains
under investigation during the past 20 years, see Ref. 3 and references therein.
However, a direct observation of the current oscillations caused by the cyclic
motion of carriers is not possible for the steady-state case.
In this paper we show that a transient {\it streaming-oscillating}
(SO) {\it photocurrent} can be excited by an ultrashort interband pump creating
electrons in the passive region after emission of the optical phonon cascade.$^4$

\begin{figure}[ht]
\begin{center}
\includegraphics[scale=0.38]{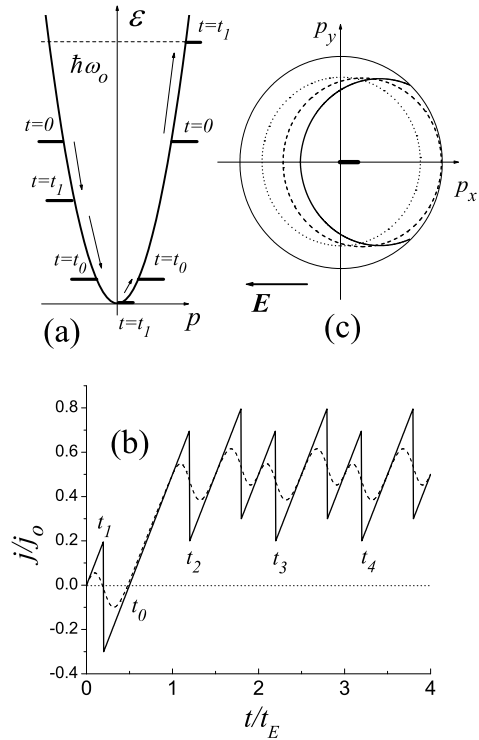}
\end{center}
\addvspace{-0.7 cm}\caption{Temporal evolution of electrons excited
at $t=0$ with $\varepsilon_{ex}=0.64$ $\hbar \omega_o$. (a) Position of 1D electrons
in the passive region at different instants in the case of $\delta$-shaped
initial distribution. (b) Corresponding transient current showing the regimes
of negative and oscillating conductivity. The dashed line is the current in the case
of a Gaussian-shaped, with the width 0.2 $\hbar \omega_o$, initial distribution.
(c) Evolution of the initial $\delta$-shaped distribution of 2D electrons:
dotted circle at $t=0$, dashed circle at $t=t_1$, and solid segment plus a
short straight line near the center (streaming component) at $t>t_1$.}
\end{figure}

Apart from the SO current, we have found that a {\it transient absolute negative
conductivity}, when the current flows in the direction opposite to the applied
field, appears if the excitation energy is close to the edge of the active region.
The phenomenon of absolute negative conductivity (ANC) of photoexcited electrons
in strong electric fields has been discussed theoretically four decades
ago$^{5,6}$ for the case of stationary excitation.
Now it is clear that the negative photoconductivity in this case cannot be
observed neither in bulk materials$^7$ nor in heterostructures,
because of accumulation of low-energy electrons with time and scattering
of high-energy electrons by the low-energy ones,$^8$ which leads to a fast
broadening of the narrow energy distribution of photoexcited electrons. The
case of ultrafast interband photoexcitation is free from these disadvantages
because the accumulation takes place only at long
times associated with energy relaxation, and the electron-electron scattering
remains ineffective if the density of excited electrons is low.
In contrast to the steady-state negative conductivity, the
effect we condider is transient, it exists in a finite time interval
after the photoexcitation. Nevertheless, the origin of this effect
is the same as in the steady-state case, and the necessary condition
for its realization is the partial inversion of the distribution of excited
electrons. Recently, a transient ANC and negative absorption of photoexcited electrons
have been suggested$^9$ for the regime of linear response, so the consideration
of the nonlinear transient response is now timely.

To describe the mechanisms of the transient ANC at $t < t_{\ss E}$ and oscillating
current at $t > t_{\ss E}$, we consider the model of one-dimensional (1D) conductor
[Fig. 1 (a,b)], where electrons are excited at $t=0$ with the momenta $\pm p_{ex}
= \pm \sqrt{2m \varepsilon_{ex}}$ corresponding to the excitation energy
$\varepsilon_{ex} < \hbar \omega_o$.
Let us assume that $p_{ex} > p_o/2$ and take into account that the electric field
accelerates the electrons according to the dynamical equation $p_t=p_{t=0} +
|e| E t$ (the electric field in Fig. 1 is assumed to be applied in the negative
direction so that the electrons move from the left to the right).
The electrons of the right group (with $p_{t=0}=p_{ex}$) come to the edge
of the active region at $t_1=(p_o-p_{ex})/|e|E$ and immediately drop to the point
$p=0$, where they do not contribute to the current. At the same instant, the
electrons of the left group (with $p_{t=0}=-p_{ex}$) have negative momentum,
$p_{t=t_1}= p_o - 2 p_{ex}$, so that the current abruptly becomes negative.
Since the momentum of the electrons of the left group increases again,
the current changes its sign from negative to positive at $t_0=t_{\ss E}/2$, when the
absolute values of the momenta for the two groups become equal. The interval of
the ANC is equal to $t_0-t_1$ and becomes maximal (equal to $t_{\ss E}/2$) when
$t_1 \rightarrow 0$, i.e. when the electrons are excited very close to the edge.
At $t > p_{ex}/|e|E$ both groups of electrons undergo cyclic motion along the right-hand
branch of the energy parabole, with the period $t_{\ss E}$. This leads to periodic
oscillations of the current with the same period [Fig. 1 (b)]. The current, expressed
in the figures here and below in units of $j_o=|e| n_{ex} p_o/m$, where $n_{ex}$ is the
excited electron density, linearly
increases with time and drops abruptly each instant when one of the groups reaches
the boundary of the active region (this occurs at $t=(k p_o \mp p_{ex})/|e| E$, where
$k$ is integer). If $p_{ex} < p_o/2$, the oscillations remain but the current is always
positive.
%If $p_{ex}=p_o/4$ or $3p_o/4$, the period of the oscillations is halved.
The main effects discussed above remain valid when the initial electron energy
distribution is a broad peak, though the sharp features of the current become smoothed.

In the 2D and 3D cases (which differ from each other only by geometrical factors),
the evolution is more complicated because the processes of optical
phonon emission returning electrons to the point ${\bf p}=0$ persist in a wide time
interval, $t_1 < t < t_1 + 2 p_{ex}/|e|E$ for the $\delta$-shaped initial distribution
$f_\varepsilon^{(ex)} \! \propto \delta(\varepsilon-\varepsilon_{ex})$. This interval
begins when the surface $|{\bf p}_t|=p_{ex}$ (here and below,
${\bf p}_t \equiv {\bf p} - e {\bf E} t$) touches the boundary of the
active region, $|{\bf p}|=p_o$, [see Fig. 1 (c)] and ends when all the excited electrons
emit optical phonons. Starting from $t_1$, the distribution in the passive
region acquires a continuous streaming part (at $t > t_1 + 2 p_{ex}/|e|E$ only
this part remains). Nevertheless, both the ANC and the transient SO current
exist in the 2D and 3D cases, and their properties are similar to those discussed
for the 1D case.

To describe the electron distribution in the passive region, $p< p_o$, we first
analytically consider the case of ballistic transport over the passive region and
then numerically calculate the current in the presence of elastic scattering
described by the collision integral $J_{el}(f|\mathbf{p} t)$. The distribution function
$f_{\mathbf{p}t}$ is governed by the kinetic equation
\begin{equation}
\left( \frac \partial {\partial t} + e\mathbf{E}\cdot \nabla _{\mathbf{p}}\right)
f_{\mathbf{p}t}= G_t \delta (\mathbf{p})+J_{el}(f| \mathbf{p} t)
%1
\end{equation}
with the initial condition $f_{\mathbf{p}t=0}= f_\varepsilon^{(ex)}$, where
$\varepsilon= \varepsilon_p \equiv p^2/2m$,$^{10}$ and the normalization
condition $n_{ex}=2\int_{V_o} d \mathbf{p}f_{\mathbf{p}t}/(2\pi \hbar )^d$, where $d=$2
or 3 for the 2D or 3D case and $V_o$ is the area or volume of the passive region.
Equation (1) assumes that the electrons instanteniously emit optical phonons so that
their penetration into the active region is neglected and the generation term on
the right-hand side, $G_t$, describes the electrons coming from the boundary of the
active region, $S_o$, directly to the point $\mathbf{p}=0$. Using the conservation of
the total number of electrons in the processes of optical phonon emission and elastic
scattering, we find $G_t=e \mathbf{E} \cdot \int_{S_o} d \mathbf{s} f_{\mathbf{p}t}$,
where the integral over the boundary $S_o$ appears owing to the relation $\int_{V_o}
d\mathbf{p\nabla }_{\mathbf{p}}f_{\mathbf{p}t}=\int_{S_o}d\mathbf{s}f_{\mathbf{p}t}$.

Even in the ballistic regime, the presence of the integral over $S_o$ in the generation
term makes Eq. (1) an integro-differential equation. One can solve this equation analytically
by separating the streaming contribution to the electron distribution function according to
$f_{\mathbf{p}t}=f^{(ex)}_{{\bf p}_t^2/2m} + g_{p_{\ss \Vert} t}  \delta({\bf p}_{\ss \perp})$,
where ${\bf p}_{\ss \perp}$ and $p_{\ss \Vert}$ are the components of the momentum perpendicular
and antiparallel to the electric field, respectively. The first part of this function
satisfies the kinetic equation (1) without the right-hand side, while the second part is
governed by the differential equation
%\vspace{-0.2 cm}
\begin{eqnarray}
\left( \frac{\partial}{\partial t} + |e| E \frac{\partial}{\partial p_{\ss \|}} \right)
g_{p_{\ss \|} t}= |e| E \delta (p_{\ss \|} )  \\
\times \left[ g_{p_o t} - {\bf e} \cdot \int_{S_o} d \mathbf{s}'
f^{(ex)}_{{\bf p}_t '^2/2m}\right],  \nonumber
%2
\end{eqnarray}
where ${\bf e}$ is the unit vector in the direction of ${\bf E}$ and the
initial condition is $g_{p_{\ss\|}t=0}=0$. Using the trajectory method, one may
transform Eq. (2) into an algebraic finite-difference equation which can be solved
at $p_{\ss\|}=p_o$. Thus, the streaming contribution takes the form
%\vspace{-0.2 cm}
\begin{eqnarray}
g_{p_{\ss \|} t}= -\theta(p_{\ss \|}) \theta(|e| E t-p_{\ss \|}) \nonumber \\
\times {\bf e} \! \cdot \! \int_{S_o} \! d \mathbf{s}' \sum_{k=0}^{k_m}
f^{(ex)}_{[{\bf p}_t ' - {\bf e}(p_{\ss \|}+k p_o)]^2/2m} ~,
%3
\end{eqnarray}
where $k_m$ is the integer part of $(|e|Et-p_{\ss \|})/p_o$. Depending on the variables
$p_{\ss \|}$ and $t$, the sum in Eq. (3) contains just one or two terms, because
$f^{(ex)}_{\varepsilon}$ is nonzero only in the region $\varepsilon < \hbar \omega_o$.

The current density, $\mathbf{j}_t=2(e/m)\int_{V_o} d\mathbf{pp}f_{\mathbf{p}%
t}/(2\pi \hbar )^d$, is written as $\mathbf{j}_t={\bf e} j_t$ with
\begin{eqnarray}
j_t=\frac{2e}{m(2\pi \hbar )^d} \biggl[ \int_{V_o} d\mathbf{p}~
{\bf e} \cdot \mathbf{p} f_{p_t^2/2m}^{(ex)} +
\int_0^{p_m} d p_{\ss \Vert} p_{\ss \Vert}  \nonumber \\
\times {\bf e} \cdot \int_{S_o} d \mathbf{s}' \sum_{k=0}^{k_m}
f^{(ex)}_{[{\bf p}_t ' - {\bf e}(p_{\ss \Vert}+ k p_o)]^2/2m} \biggr],
%4
\end{eqnarray}
where we have used Eq. (3) and denoted $p_m={\rm min}(p_o, |e|Et)$. Analytical
integration in Eq. (4) is possible for the $\delta$-shaped initial distribution,
$f^{(ex)}_{\varepsilon}=[n_{ex}/\rho_{dD}(\varepsilon_{ex})] \delta(\varepsilon-
\varepsilon_{ex})$, where $\rho_{dD}(\varepsilon_{ex})$ is the $d$-dimensional
density of states at the excitation energy.
If $|e|Et < p_o-p_{ex}$, the current linearly increases with time,
$j_t=e^2 n_{ex} E t/m$, while at $|e|Et > p_o-p_{ex}$ its temporal dependence
becomes complicated.
%This dependence can be presented as $j_t= (|e| n_{ex}/m)(P_{1t} +P_{2t})$.
%The negative response at small times is caused by the first part of $j_t$, while
%the second part, originating from the streaming contribution to the distribution
%function, is always positive. At $|e|Et > p_o+p_{ex}$, when only the
%latter contribution remains, the current becomes periodic.

\begin{figure}[ht]
\begin{center}
\includegraphics[scale=0.48]{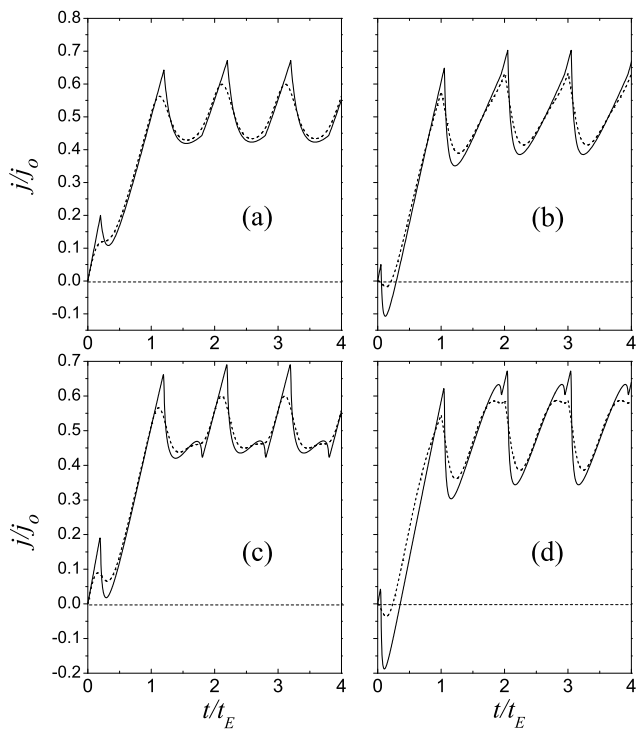}
\end{center}
\addvspace{-0.7 cm}\caption{Transient current in 3D (a,b) and 2D (c,d) conductors
for $\delta$-shaped initial photoexcited distribution (solid) and for a Gaussian
peak $F^{(ex)}_{\varepsilon} \! \propto \exp[ -(\varepsilon-\varepsilon_{ex})^2/\Delta^2]$
(see Ref. 10) with the width $\Delta =$0.2 $\hbar \omega_o$ (dashed). The excitation
energies are $\varepsilon_{ex}=0.64$ $\hbar \omega_o$ (a,c) and 0.9 $\hbar \omega_o$ (b,d).}
\end{figure}

The results of numerical calculations of the current according to Eq. (4)
for the 3D and 2D cases are shown in Fig. 2. We have chosen two excitation
energies, close to the edge of the active region and far from this edge, and
plot the temporal dependence of the current both for infinitely narrow
($\delta$-shaped) distribution of excited electrons and for a wide,
Gaussian-shaped distribution. The transient current at large $t$ oscillates
around the value of the steady-state streaming current, $j_o/2$.
The plots for the $\delta$-shaped
excitation show sharp (kink-like) features at $t=(k p_o \mp p_{ex})/|e|E$,
whose origin is explained in the above discussion of the 1D case (Fig. 1).
The plots for the wide excitation close to the edge show similar sharp
features at $t=k t_{\ss E}$, which are caused by the abrupt drop of the
electron distribution at $p=p_o$ (these features are absent if the
electrons are excited far from the edge). For the excitation close
to the edge, there exist regions of ANC. One can find a simple
criterion describing the existence of ANC at small times: $j_{t \rightarrow 0} < 0$
at $f^{(ex)}_{\hbar \omega_o} > n_{ex} [d \hbar \omega_o \rho_{dD}(\hbar \omega_o)]^{-1}$,
where $d=$2 or 3 is the dimensionality of the electron gas. In other words, the ANC
at small times always takes place if the number of electrons excited
close to the edge of the active region is sufficiently large. Comparing
the plots (b) and (d) in Fig. 2, one can see that the absolute value of the
negative response is larger for 2D electrons.

The elastic scattering tends to isotropise the distribution function of electrons.
For this reason, it is important to study the effect of this scattering on both
the ANC and SO current. To do this, we consider Eq. (1) with the collision integral
on the right-hand side. We restrict ourself by the 2D case, when the energy-independent
scattering rate, $\nu$, can be used and this integral acquires the most simple form
$J_{el}(f| \mathbf{p} t)= -\nu (f_{\mathbf{p}t} - f^{(s)}_{\varepsilon t})$, where
$f^{(s)}_{\varepsilon t}$ (with $\varepsilon=\varepsilon_p \equiv p^2/2m$)
is the isotropic part of $f_{\mathbf{p}t}$, which is obtained
by averaging $f_{\mathbf{p}t}$ over the angle of the vector ${\bf p}$.
Then Eq. (1) is reduced to the integral equation
%\begin{eqnarray}
%f_{\mathbf{p}t} = e^{-\nu t} f^{(ex)}_{{\bf p}_t^2/2m}
%+ (|e|E)^{-1} \theta(p_{\Vert}) \theta(|e|Et-p_{\Vert}) \delta({\bf p}_{\bot})
%\nonumber \\
%\times e^{-\nu p_{\Vert}/|e|E} G_{t-p_{\Vert}/|e|E}
%+ \nu \int_0^t dt' e^{-\nu(t-t')} f^{(s)}_{{\bf p}_{t-t'}^2/2m ~ \! t'}~,
%%5
%\end{eqnarray}
\begin{eqnarray}
f_{\mathbf{p}t} = e^{-\nu t} f^{(ex)}_{{\bf p}_t^2/2m} \!
-  \theta(p_{\ss \Vert}) \theta(|e|Et \! - \! p_{\ss \Vert}) \delta({\bf p}_{\ss \bot})
e^{-\nu p_{\Vert}/|e|E}
\nonumber \\
\times  {\bf e} \!  \cdot \!  \int_{S_o} \! \! d \mathbf{s}' f_{{\bf p}'  t-p_{\ss \Vert}/|e|E}
+ \nu \int_0^t dt' e^{-\nu(t-t')} f^{(s)}_{{\bf p}_{t-t'}^2/2m ~ \! t'}
%5
\end{eqnarray}
which can be transformed to a closed integral equation for $f^{(s)}_{\varepsilon t}$.
We have solved this equation numerically, by iterations. The results presented in Fig. 3 (a)
show that the scattering suppresses both the magnitude and time interval of the ANC, while
Fig. 3 (b) demonstrates how the amplitude of the oscillations decreases with
time and with increasing scattering. An essential point is that the suppression
of the ANC is not so strong as one may expect, especially as concerns the time
interval. If the scattering rate is $\nu=$50 $t_{\ss E}^{-1}$, the ANC interval decreases
only by half in comparison to the ballistic case and becomes approximately 8 times
greater than $\nu^{-1}$. However, the scattering considerably suppresses the absolute
value of the current. On the other hand, the suppression of the SO current is already
strong at $\nu \simeq t_{\ss E}^{-1}$, and the oscillations become washed out as $\nu$
increases.

\begin{figure}[ht]
\begin{center}
\includegraphics[scale=0.42]{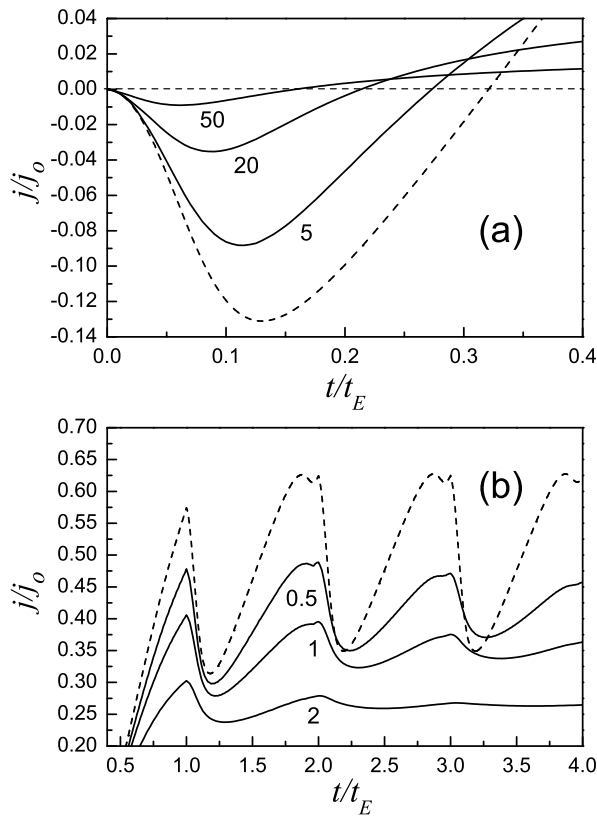}
\end{center}
\addvspace{-0.7 cm}\caption{Transient current in 2D conductors in the presence
of elastic scattering in the ANC (a) and SO (b) temporal regions. The initially
excited distribution is the Gaussian of width $\Delta=$0.1 $\hbar \omega_o$ centered at
$\varepsilon_{ex}=$0.9 $\hbar \omega_o$. The dashed curves show the ballistic case
($\nu=0$), while the solid curves are marked with the corresponding values of
$\nu t_{\ss E}$.}
\end{figure}

Let us discuss the results in terms of numerical parameters. The ballistic
expression (4) is suitable for describing the transient response at the times
smaller than the collision times. In clean samples, it is limited by the acoustic-phonon
scattering rate $\nu_{ac}$, according to $t < \nu_{ac}^{-1}$. The measured low-temperature
collision times for 2D electrons in clean GaAs-based structures$^{11}$ are of about 150 ps.
As follows from Fig. 2 (d), the region of ANC terminates at $t \simeq 0.35$ $t_{\ss E}$.
Therefore, the influence of the scattering on the ANC is not essential at $E > 0.35$
$\nu_{ac} p_o/|e|$, which gives us an estimate $E >4$ V/cm for GaAs quantum wells.
The numerical calculations according to Eq. (5) show that the ANC can be as well observed
in the fields smaller by one order of magnitude, since the elastic scattering does not
lead to a strong suppression of the effect; the time interval where the ANC exists in this
case extends over a nanosecond scale. In order to generate several periods of the
transient oscillations in the streaming redime, the electric field should be considerably
higher, $E >$ 10 V/cm. In bulk semiconductors,$^1$ $\nu_{ac}^{-1}$ at low temperatures
exceeds 100 ps (moreover, for InSb $\nu_{ac}^{-1}>1$ ns) so that the estimates given above
remain valid. Since the frequency of the SO current, $2 \pi/t_{\ss E}$, is proportional
to $E$, one can obtain high-frequency oscillations by increasing $E$ up to $p_o \nu_{o}/|e|$.
In particular, application of high fields, $E \simeq 3 \times$10$^5$ V/cm, to wide-gap
materials such as GaN (where $p_o$ and $\nu_o$ are large) can generate oscillations of
about 0.5 THz, while for GaAs the attainable frequencies are several times smaller.

Now we discuss the main assumptions. The neglect of inelastic scattering in the passive
region assumes that the energy given to electrons via their acceleration by the field is
much higher than the energy which they emit or absorb when scattering by acoustic phonons.
This approximation is well justified even at $t \gg \nu_{ac}^{-1}$ because of smallness
of the sound velocity $s$ and remains valid under the condition $E \gg \nu_{ac} m s/|e|$.
The neglect of a finite energy of electron penetration into the active region
is justified if this energy is small in comparison to $\hbar \omega_o$. This requires
$\nu_o t_{\ss E} \gg 1$, which is easily attainable since the typical times of optical
phonon emission in quantum wells and bulk materials do not exceed 0.4 ps. Next,
since we consider the interaction of electrons with dispersionless optical phonons,
the presented results have a limited applicability for heterostructures.
Finally, the neglect of electron-electron scattering can be justified if the excited
electron density $n_{ex}$ is small enough so that this scattering mechanism is less
important than the elastic scattering. Comparing the elastic scattering time of the
order of 100 ps to the electron-electron scattering time,$^1$ one can find
$n_{ex}< 10^{14}$ cm$^{-3}$ for bulk GaAs and $n_{ex}< 10^{8}$ cm$^{-2}$
for GaAs quantum wells.

In conclusion, we have proposed and theoretically verified a mechanism of
the transient absolute negative conductivity, which is related to the process of
transformation of the nearly isotropic initial electron distribution to the strongly
anisotropic (streaming) one. This effect appears to be robust against the elastic scattering,
can exist over one nanosecond, and can be investigated by standard experimental methods.
We have also described transient oscillations of the electric current caused by
the streaming effect. In strong electric fields, the frequencies of these
oscillations belong to the THz spectral region.


\begin{thebibliography}{11}
\bibitem{10}
If the initial excitation creates electrons both in the passive and active
regions, with the distribution function
$F^{(ex)}_{\varepsilon}$, we have $f^{(ex)}_{\varepsilon}=\theta(\hbar \omega_o-
\varepsilon) \sum_{k=0}^{\infty} F^{(ex)}_{\varepsilon + k \hbar \omega_o}$, where
the terms with $k \neq 0$ describe the electrons rapidly transferred to
the passive region via optical phonon emission.
Instead of using the initial condition, one may add a generation term $\delta (t)
f^{(ex)}_{\varepsilon_p}$ describing instantenious filling of the passive region
to the right-hand side of Eq. (1).
\end{thebibliography}
\end{document}